\documentclass{elsart}
\usepackage{epsfig}
\journal{Physica D}
\begin{document}
\begin{frontmatter}

\title{A Trade-Investment Model for Distribution of Wealth}
\author[South]{Nicola Scafetta},
\corauth[cor]{Corresponding author.}
\ead{ns2002@duke.edu}
\author[Oxford]{Bruce J. West $^{a,}$}
\author[South]{and Sergio Picozzi}

\address[South]{Department of Physics and Free Electron Laser Laboratory, Duke University, Durham, NC 27708, USA}

\address[Oxford]{Mathematics Division, Army Research Office, Research Triangle Park, NC 27709-2211, USA }

\begin{abstract}
Econophysics provides a strategy for understanding the potential mechanisms
underlying   the anomalous distribution of wealth found in real societies. We present a 
computational nonlinear stochastic model for the distribution of wealth that depends upon three parameters and two mechanisms: trade and investment. To avoid economic paradoxes, the trade mechanism is assumed to be
 related to the poorer trader's wealth and to  statistically
advantage the poorer of the two traders.  The two mechanisms together are shown to generate a
distribution that reproduces the full range of the empirical wealth distribution, and not only the inverse power-law tail that Pareto found in western societies at the end of the 19th century.
\end{abstract}


%Keywords: Anomalous, condensation, interaction, Pareto, wealth.

%Pacs: 89.65.Gh, 87.23.Ge, 05.45.Df, 02.60.Cb

\end{frontmatter}

\section{Introduction}

The science of complexity leads to an understanding of
complex physical phenomena through the construction of both equilibrium and
dynamical models. These ideas have found application in engineering,
biology, sociology and economics; specifically, in areas where phenomena
have both the characteristics of randomness and determinism \cite{west1}, present nonlinear and non-extensive behaviors and/or anomalous distributions. To understand the main mechanisms that lead to the empirical distribution of wealth in a society is one of such applications.

More than a century ago the Italian
sociologist and economist, Vilfredo Pareto determined that the
cumulative probability of high income in western societies obeys to an inverse
power law, 
 \begin{equation}\label{parretolaw}
P(w) = \int _{w}^{\infty }p(x)~dx~\propto \frac{1}{w^{\mu}}~, 
\end{equation}
with $1 \leq \mu \leq 2$ \cite{pareto} and where $p(x)$ is the probability density function (pdf) of wealth.  

 It is straightforward to observe that {\it wealth} and {\it income} are different concepts \cite{black}. However, even if individual wealth and income are   related in a very complex way, their empirical distributions, regarding an entire society, look similar. Compare, for example, the shape of the wealth distribution for United Kingdom studied in Ref. \cite{dragulesco3} with that of the income distributions for United Kingdom and other countries studied in Refs. \cite{dragulesco3,souma}. Fig. 1 shows the cumulative distribution of wealth and income data for United Kingdom, downloaded from Inland Revenue, the British tax agency \cite{stati}. In particular, the reconstructed wealth distribution of the whole UK population was obtained by using an adjustment procedure applied to the data of all assets and liabilities of a person that must be reported at his or her death for the purpose of inheritance tax \cite{dragulesco3}.  

Since
Pareto's time a new field of research, \textit{econophysics}, has
emerged showing that physically-based models yield possible
explanations of economic mechanisms in society \cite{stanley2}.
In 1983 Montroll and Shlesinger \cite{montroll2} 
demonstrated that renormalization group scaling provides an
interpretive context for Pareto's inverse power-law distribution. In fact, 
according to the renormalization group theory, a function $F\left( x\right) $ is homogeneous if there are two
constants $a$ and $b$, such that $F(x) =a~F(bx)$. The simplest
 solution to this equation is given by a Pareto law (\ref{parretolaw}) with $\mu =\log a/\log b$ for $0<(a,b)<1$,
\cite{stanley1}.  Moreover,
Montroll and Shlesinger argue that if
$g\left( w\right) $ is the distribution of wealth excluding the
very rich, the total distribution should have the anomalous form
\begin{equation}
G\left( w \right) =g \left( w \right) +a~G\left( b~w \right)
\label{renorm1}
\end{equation}
where $b$ is a mechanism that increases the variance of the distribution and
$a$ is the probability of that mechanism being present in the society.
For very large $w$ the distribution obtained from solving (\ref{renorm1}) is
a Pareto's law
\begin{equation}
\lim _{w \rightarrow \infty }G ( w ) \propto  \frac{1}{w^{\mu }},
\label{renorm2}
\end{equation}
and for low-medium  level of wealth the distribution is
\begin{equation}
\lim _{w \leftarrow \infty }G\left( w\right) =g\left( w\right) .
\label{renorm3}
\end{equation}
Thus, they explain the inverse power-law tail in the distribution of wealth as
a scaling effect in an equilibrium statistical society. On the other side, Eq. (\ref{renorm1}) suggests that the distribution of wealth at high and low ends is affected by different mechanisms. Indeed, full range empirical distributions of wealth or income \cite{dragulesco3,souma} do present such a differentiation. For example, Fig. 1 shows the cumulative probability of wealth during 1996 and of income during 1998-1999 in United Kingdom  \cite{stati}. It is easy to see the separation between low-medium level and high level that is characterized by a Pareto law and involves almost 1\% of the population.  In this paper we present   a nonlinear physically-based model for wealth distribution able to interpret it.

While many  have used variants of the Montroll-Shlesinger argument to explain 
several phenomena  \cite{list},
 here, we take a different
approach and identify possible dynamical and stochastic mechanisms that lead to the observed distribution of wealth. 
 We draw inspiration
from the evolution equation postulated by Bouchaud and M\'{e}zard
\cite{bouchaud1} that supposes distribution of wealth depends  on trades and investments, but we introduce different  rules for trading dynamics. In Sec. 2 we review some models found in the econophysics  literature like the mean-field approximation of the Bouchaud and M\'{e}zard
model \cite{bouchaud1} and the kinetic theory approximation \cite{dragulesco3,dragulesco1,dragulesco2,Chakraborti} and explore some economic paradoxes induced by these approximations about the trade mechanism. In Sec. 3 we
postulate a nonlinear stochastic trade-investment model that does not encounter the
difficulties of the above approximations and  we determine the
properties of the model through numerical solutions. 
In
Sec. 4 we argue the validity of  our trade-investment model in interpreting the full range (both high and low ends) of the empirical wealth distribution as that recently found in the United Kingdom and shown in Fig. 1. Finally, we draw conclusions in Sec. 5.

\section{Economic paradoxes of some asset exchange linear models}

Distribution of wealth depends mainly on two
mechanisms: trade and investment \cite{dragulesco3,souma,bouchaud1,dragulesco1,dragulesco2,Chakraborti,ispolatov,malcai,burda}. We stress that in the physically-based  models {\it trade and investment} should always be understood in a  generalized sense. A linear model proposed by Bouchaud and Mezard \cite{bouchaud1}, borrowed from the
physics of directed polymers,  describes the dynamics of the individual wealth $W_{i}(t)$ in a given a society of $N$ agents  by mean of the following equation 
\begin{equation}
\frac{dW_{i}}{dt}=\eta _{i}(t)~W_{i}+\sum_{j=1(\neq
i)}^{N}J_{ij}~W_{j}-\sum_{j=1(\neq i)}^{N}J_{ji}~W_{i}~.  \label{firsteq}
\end{equation}
The component $\eta _{i}(t)~W_{i}$ is a Gaussian multiplicative process with variance $\sigma$ that
simulates the investment dynamics and it is related, for example, to the temporal
change in the value of stocks \cite{bouchaud1,levy,Boucha,sornette,solomon}.  The two sum terms of Eq. (%
\ref{firsteq}) describe the trade interaction network between the agent $i$
and all other agents in the society and $J_{ij}$ is the exchange
rate between agents i and j.

Usually, Eq. (\ref{firsteq}) is solved with the simplifying assumption that the
exchange rate for all agents is the same, i.e., $J_{ij}\equiv J/N$, so that
Eq. (\ref{firsteq}) reduces to
\begin{equation}
\frac{dW_{i}}{dt}=\eta _{i}(t)~W_{i}+J(\overline{W}-W_{i})~,
\label{firsteq2}
\end{equation}
where $\overline{W}=N^{-1}\sum_{i}W_{i}$ is the mean wealth. Eq. (\ref
{firsteq2}) is the \textit{mean-field} approximation of the trade process
\cite{bouchaud1}. Such an approximation is useful because Eq. (\ref{firsteq2}) can be associated with a solvable Fokker-Planck equation with the
following equilibrium pdf solution
\begin{equation}\label{probden1}
p_{eq}(w)=\Psi \exp\left[\frac{1-\mu}{w}
\right]~\frac{1}{w^{1+\mu}}~,
\end{equation}
where $\Psi =(\mu -1)^{\mu }/\Gamma [\mu ]$ is the normalization
constant and $\mu =1+J/\sigma ^{2}$ is the Pareto exponent
\cite{bouchaud1}.

However, the power-law tail of Eq. (\ref{probden1}) is due to the multiplicative term in (\ref{firsteq2}) \cite{lindenberg}, and does not, by itself alone, assure
the goodness of the mean-field approximation in describing a realistic trade
process. In the absence of the multiplicative process, the mean-field
approximation  causes the wealth of all economic agents
to exponentially converge toward the mean wealth $\overline{W}$. In fact, the
solution of Eq. (\ref{firsteq2}) without the multiplicative process is
\begin{equation}
W_{i}(t)=\overline{W}+(W_{i}(0)-\overline{W})\exp [-J~t]~,  \label{esexptg}
\end{equation}
implying that the trade dynamics has the effect of equalizing the wealth
among all members of the society so that the observed condensation of
the wealth is interpreted as due solely to the presence of the investment
process.

This form of trade, implicit in the mean-field
approximation, does not seem realistic because it assumes that the agent $i$ gives the
fraction $J/N$ of his own wealth to the agent $j$ and, in exchange,
receives the same fraction $J/N$ of agent $j$'s wealth. So, a
consistent amount of wealth, proportional to the rich agent's wealth, passes
to the poor; an outcome that is not encountered in the real world. Instead,
in a real trade, the rich always risk
less than do the poor (in a statistical sense), because of their greater
resources. This property should be fulfilled by the trade mechanism such that, in addition to the investment mechanism, the trade, too,  has
the capacity to produce a condensation of wealth. 

A similar economic paradox is shared by the kinetic theory approximation
when applied to the trade interaction between two agents \cite
{dragulesco3,dragulesco1,dragulesco2,Chakraborti}. This approximation yields a
Maxwell-Boltzmann exponential distribution for energy ($p(E)\propto e^{-E/kT}$).  In fact, the kinetic
theory approximation assumes that the transactions of wealth  are similar to the transaction of energy in random elastic collisions between particles.   Therefore, according to  such an analogy,  in a trade transaction the total
wealth-energy (or a fraction of it according to the reaction scheme $
[W_{i},W_{j}]\rightarrow [\gamma W_{i}+\varepsilon (1-\gamma
)(W_{i}+W_{j}),\gamma W_{j}+(1-\varepsilon )(1-\gamma )(W_{i}+W_{j})]$, \cite
{Chakraborti}) of both traders should mix, become randomly divided between
the two traders and, on average, leave the two of them with the same amount
of wealth. Also in this case, in a trade interaction between the rich and
poor, the poor have an extraordinary chance to significantly increase their wealth;
an unrealistic outcome. Moreover, the empirical distributions of wealth or income are not monotonic \cite{stati}; they increase, reach a maximum and, finally, decrease
as an inverse power-law function. Therefore,  the empirical distribution of wealth or income can not be recovered by the monotonically decreasing exponential distribution of Maxwell and Boltzmann.

It is also not convincing that in a trade a randomly
selected agent loses a fixed fraction of his own wealth to a randomly
selected winner \cite{dragulesco1,ispolatov} according to the reaction
scheme $[W_{i},W_{j}]\rightarrow [W_{i}-\gamma W_{i},W_{j}+\gamma W_{i}]$,
where $0<\gamma <1$ is the fraction of wealth-energy of the loser gained by
the winner. Such a trade mechanism would imply that in a transaction the
richer the trader, the more he/she may lose in favor of the poorer trader.

These economic paradoxes, common to both the mean-field and kinetic theory
approximations, seem to be related to the fact that those linear models  assume a type of symmetric status of the two traders.
Both  approximations assume that the amount of
wealth that can move from one agent to the other may be linearly related to the
wealth of both agents or, at least, to the wealth of one of them randomly
selected. We make what we believe to be a more realistic assumption
regarding the common trade interaction dynamics.

\section{A nonlinear trade-investment model}

We  assume certain characteristics for the trading transaction that avoid
the above paradoxes. First, we assume that in a trade between the rich and poor,
the wealth that  moves from one agent to the other is  related  to the
wealth of the poorer of the two traders. In fact, only in a robbery the poorer trader can get a fraction of the wealth of the richer trader and the distribution of wealth of an entire society cannot  reasonable be  a product of  a  ``theft-and-fraud" mechanism.  Second, the trade mechanism has to take in account the role played by the
prices in mediating exchange and how these prices emerge from ideal
negotiations among agents that may belong to different social classes. Finally, we have to consider that a transaction of wealth in a trade is related to the difference between  the price and the  value of the asset.

We  modify the linear Eq. (\ref{firsteq}) and suppose that the i-th agent's
wealth $W_{i}(t)$ evolves
according to the  stochastic nonlinear equation
\begin{equation}
W_{i}(t+1) = W_{i}(t) + r_{i}\xi(t) W_{i}(t)+\sum_{j=1\left( \neq i\right) }^{N}w_{ij}(t).
\label{firsteq6}
\end{equation}
Here, as in Eq. (\ref{firsteq}), the component $r_{i}~\xi ~W_{i}$ is a Gaussian
multiplicative process that simulates investment dynamics. The variable $\xi(t)$ is a Gaussian random variable and the standard
deviation $r_{i}>0$ is the \textit{individual investment index}. We
 interpreted  $r_{i}=V\Pi
_{i}$ where $V$ is the \textit{global investment index}  of the entire society, and $\Pi
_{i} $ is the percentage of wealth $W_{i}$ that the $i$-th agent actually
invests. The summation in Eq. (\ref{firsteq6}) describes the trade
transactions between agent $i$ and all other agents. If in the temporal interval $[t:t+1]$ two particular agents do not actually trade, there is no transfer of wealth between them, that is, $w_{ij}(t)=0$. 

In a trade
there is a flow of wealth between the two agents only if, in the case
of the barter, the value of the two exchanged assets is different, or, in
the case of the purchase, the value of the asset is different from the price
paid for it. By expressing this concept in an equation, if the trader $i$
is the seller and the trader $j$ is the buyer, the exchanged wealth quantity $w_{ij}$ is
given by 
\begin{equation}
w_{ij}=price_{asset}-value_{asset}~.  \label{prival}
\end{equation}
In fact, if  the price and the value of the asset coincide, the
trade would produce only a transfer of items and money from one agent to
the other,
but there would not be any transfer of wealth. Instead, if the seller (buyer)
succeeds in selling (buying) an item for a price that is higher (lower) than
the actual value of the item, the seller (buyer) gains wealth from the buyer
(seller). 
Therefore, the trade interaction between two agents is assumed to follow the scheme
$[W_{i}(t),W_{j}(t)]\rightarrow [W_{i}(t+\delta t),W_{j}(t+\delta t)]$,
where
\begin{equation}
[W_{i}(t+\delta t),W_{j}(t+\delta t)] = [W_{i}(t)+w_{ij}(t),W_{j}(t)+w_{ji}(t)]~.  \label{varfoo}
\end{equation}
Because of Eq. (\ref{prival}), the trade variable $w_{ij}$ is antisymmetric  ($w_{ij}=-w_{ji}$), may be positive
or negative and indicates the amount of wealth that moves from one agent to
the other in each trade.  The antisymmetric nature of the trade variable $w_{ij}$ implies that the
trade itself, contrary to the investment process, conserves the total 
wealth because it  can only move wealth from one agent to another, but can neither create nor destroy wealth. 

The amount of wealth $w_{ij}$ that may move from the trader $i$ to the
trader $j$ is a stochastic nonlinear variable that  has to be a fraction of the wealth of the poorer of
the two agents and depends nonlinearly on the wealths of both traders. The simplest hypothesis consistent with these assumptions is
to suppose that $w_{ij}$ is a Gaussian variable with density
\begin{equation}
p(w_{ij})=\frac{1}{\sigma ~\sqrt{2\pi }}~\exp \left[ -\frac{(w_{ij}-%
\overline{w}_{ij})^{2}}{2\sigma ^{2}}\right] ~,  \label{diswea}
\end{equation}
where $\overline{w}_{ij}$ is the mean wealth that may move between the
two traders $i$ and $j$, 
\begin{equation}
\sigma =h~W_{ij}
\end{equation}\label{min55}
 is the standard deviation of the
distribution of $w_{ij}$ where $0<h<1$ indicates the fraction of $W_{ij}$
that may be involved in the trade, and the quantity
\begin{equation}
W_{ij}=W_{ji}=\min (W_{i},W_{j})~  \label{minggl}
\end{equation}
is the lesser of the two agents' wealth. We refer to the index $h>0$ as the \textit{%
poverty index}. In fact, in a poor society the ratio of cost to wealth is higher and,
therefore, the higher is the amount of wealth that may move in a single-trade transaction. The choice of the expression that indicates  the  mean wealth $\overline{w}_{ij}$ that may move from the trader $i$ to the trader $j$ is of crucial importance. For reasons explained subsequently, we assume that  
\begin{equation}
\overline{w}_{ij}=\alpha _{ij}~h~W_{ij}~  \label{diswe9a}
\end{equation}
where the variable $\alpha _{ij}$ is given by a nonlinear expression depending on the wealths of both traders, for example,
\begin{equation}
\alpha _{ij}=f~\frac{W_{j}-W_{i}}{W_{j}+W_{i}}~.  \label{alpha}
\end{equation}
We refer to the index  $f> 0$ as  the \textit{social index} because it  provides an advantage to
the poorer of the two traders. 
 In fact, by using Eqs. (\ref{diswea}), (\ref
{diswe9a}) and (\ref{alpha}) it is easy to prove that if the wealths $W_{i}$ and $W_{j}$ are
almost the same, then $\alpha _{ij}\approx 0$ and both traders have an equal
chance of doing either a good or bad deal. If, instead, for example, 
$W_{j}\gg W_{i}$, we have $\alpha _{ij}\approx f$, the distribution 
$p(w_{ij})$ is shifted toward positive values and the trader $i$, 
that in this example would be the poorer, has a better
chance to do a good deal and increase his own wealth. Finally, the amount of moving wealth is always a fraction of the wealth of the poorer of the two agents.

A possible interpretation of the trade interaction mechanisms described by Eqs. (\ref{prival}), (\ref{varfoo}) and (\ref{diswea}) may be formulated by considering that in a trade
 there is a transfer of wealth
between two agents only if the value of the asset is different from the
price paid for it, Eq. (\ref{prival}). The  variable $w_{ij}$ of Eqs. (\ref{prival}), (\ref{varfoo}) and (\ref{diswea}) measures such a difference and is stochastic because  the price of an asset, or similar assets, is not unique  but varies within a range in a real society.   So, there may be the possibility to buy or sell an asset at a price that may be lower or higher than the value of that asset. We observe that this  price dispersion  is due to the fact that the price is an agreement between the two traders that follows an
explicit or implicit trade negotiation and different negotiations may yield different prices for the same item. Therefore, we are not assuming the {\it ``law of
one price"}  \cite{lawoneprice}, which is a simplistic  assumption made in most economic
models (including Pareto's \cite{mauro}) that attempt to dynamically address
the problem. On the contrary,  the value of an item is a
concept that involves the entire society and the total wealth of an individual is measured by the total {\it value} of his or her belonging and not by the {\it price} of such possessions.

The stochastic nonlinear bias, see Eqs. (\ref{diswe9a}) and (\ref{alpha}), that favors the poorer trader is due to the fact that a trade can take place only if the two agents, through 
negotiation, reach an agreement about the price of the asset. The trade
transaction has a higher probability to occur if the price is below a
threshold at which the buyer would like to buy. Because this threshold
increases with the total wealth of an agent, when a wealthy agent would like
to buy something from a poorer agent, there is a higher probability that the
transaction occurs at a higher price than when a wealthy agent would like to
sell the same item to a poorer agent. Alternatively, we may say that the poor are constrained by their poverty to be more careful in their trades
and therefore, for example, they may often look for the best price
opportunity for saving money. On the other hand, because of their economic
strength, the rich may be willing to pay a premium to purchase items. This asymmetric disadvantage-advantage
tends to disappear when the two traders are economically equivalently. 

In the next subsections we investigate step by step the properties of the trade-investment model. Subsections 3.1 and 3.2 are devoted to study the properties of the trade interaction without any investment.

\subsection{Symmetric-chance trade-alone model}

Let us assume $h> 0$,  $f=0$ and $r=0$, that is, we have a trade-alone economy according to which  both agents have the same chance to gain or lose and in which there are no investments.  We observe that $f=0$ implies $\overline{w}_{ij}=0$ in Eq. (\ref{diswea}), therefore, the distribution $p(w_{ij})$  (\ref{diswea}) is symmetric and centered on  zero.    
 In this model it is easy to prove that
almost all the wealth of our ideal society  concentrates in the
hands of a few agents. In fact, because the wealth $w_{ij}$ is related to
the poorer agent's wealth, through the standard deviation $\sigma=h ~W_{ij}$ of the probability distribution (\ref{diswea}), 
the risk for the rich trader is smaller because if he/she loses a certain amount
of wealth in a trade-transaction, the loss is a smaller fraction of his/her
own wealth than that which the poorer agent may lose. Consequently, there is
a high probability in this model that few people accumulate almost the
entire wealth available and the others become devastatingly poor.

Fig. 2 shows the huge wealth gap between the rich and poor produced by the
symmetric-chance model.  This
 gap increases with the number of trade interactions and the process is boosted by
increasing the poverty index $h$. In fact, in such an eventuality  the poor may lose a larger portion of their
wealth in each single transaction. If the model is implemented with a wealth
threshold below which an agent is considered economically impotent, the entire
wealth of the society will concentrate in the hands of one person while all the others die.

This economic
catastrophe, the \textit{gambler's ruin} problem \cite{feller}, is due simply to the intrinsic economic strength of the rich
and, therefore, such a collapse may happen also  without any economic abuse or
exploitation of the poor by the rich. This 
condensation aspect of the trade-interaction dynamics is completely masked
by the mean-field approximation of the Bouchaud and Mezard's model \cite{bouchaud1} that, without any investment, yields a uniform
distribution of wealth. Moreover, the strong wealth condensation of the fair
trade economy also explains why, to avoid economic collapse, in real societies
there exist a number of  mechanisms that
have the effect of redistributing wealth  by
advantaging the poor.

\subsection{Asymmetric-chance trade-alone model}

The redistribution mechanism that
advantages the poor is implemented by supposing that the probability
distribution of transaction of wealth $w_{ij}$, Eq. (\ref{diswea}), is biased
in favor of the poorer trader. The bias is introduced through the mean
transaction wealth $\overline{w}_{ij}\neq 0$ because we now assume 
the social index $f>0$ (\ref{alpha}), and we again assume $h>0$ and $r=0$. A social index $f>0$ provides an advantage to
the poorer of the two traders. 

The asymmetric-chance model redistributes the wealth, as shown in Fig. 3, and
leads to a stable distribution. In addition, this model shows the
emergence of a large middle class, followed by a smaller poor class and an
even smaller rich class. The economic gap between the richest and poorest is not
unrealistically wide, as it was in the symmetric-chance model.  The condensation of wealth, that is, the wealth separation
between the richest and poorest increases by decreasing the social index $f$.
Fig. 3 also shows that this type of distribution of wealth can be well
fitted with a Gamma-like distribution
\begin{equation}
p(w)=a~(w-c)^{\eta }~\exp (-d~w),  \label{fitt1}
\end{equation}
that has an exponential-like tail. In the computer simulation  we suppose a society of 100,000 economical agents with an initial uniform distribution of wealth and determine the wealth pdf after 100 million random trade-interactions between two  randomly selected agents.

Fig. 4 shows how the asymmetric-chance model depends on the poverty index $h$.
By keeping the social index $f$ fixed, the wealth gap between the richest
and poorest increases by increasing the poverty index $h$. In fact, in a
rich society, one based on trade alone, wealth is better distributed because it is more difficult for wealth to
condense into the hands of a very few
 agents because people need less money to buy items. Consequently, in each
trade-interaction agents may lose a smaller fraction of their wealth.

In summary, our trade interaction model is based on the balance of two
opposing contributions: (i) the condensation mechanism of the  fair trade dynamics; (ii) a redistribution factor giving a stochastic
advantage to the poor. When both processes are present, the trade
interaction mechanism produces a stable distribution of wealth well fitted with a Gamma-like distribution. 

\subsection{Pareto tail in the trade-investment economy}

In the last step we assume  all three indexes  of the trade-investment model are positive: $h> 0$,  $f>0$ and $r>0$.
 Fig. 5 shows the distribution of wealth for three computer simulations in which we implement the previous model by assuming that every 10,000 trades the wealth of all agents is re-initialized by the multiplicative process.  We fit the tail of
the computer-generated wealth pdf with a Pareto distribution $1/x^{\delta },$
where $\mu =\delta -1$ is the Pareto's exponent. These computer-generated wealth pdfs may be {\it apparently} well-fitted by
functions of the form
\begin{equation}
p(w)=a~w^{\gamma }/(1+b~w)^{\gamma +\delta }~.  \label{fitt2}
\end{equation}
 However, we stress that in the
absence of the analytic solution to Eq. (\ref{firsteq6}), we do not know the
exact analytic form for these computational distributions and, as we will explain
in Sec. IV, the true solution to Eq. (\ref{firsteq6}) is more complex than
 Eq. (\ref{fitt2}). However, we notice that
the functions (\ref{fitt1}) and (\ref{fitt2}) may be related, one the other,
in the limit $r\to 0$ because an exponential function can be obtained from a power-law function via an appropriate limit procedure \cite{tsallis}.

Fig. 5 shows  that wealth condensation increases by increasing the
investment index $r$ and by decreasing the social index $f$. The connection
with the poverty index $h$ is more complicated and will be explained in the next
section. As in the asymmetric-chance model, we obtain a stable distribution
of wealth $p(w)$ that partitions society into three classes.  Fig. 6 shows the cumulative probability $P(w)$
for an economy with $f=0.3$, $h=0.05$ and $r=0.075$, fitted with the Pareto
distribution having an index $\alpha =\delta -1=1.5$. This figure is
consistent with the fit to U.S. income data made in Figure 35 of Montroll
and Badger \cite{montroll}. According to a more recent study made by Dr\u{a}gulescu
and Yakovenko \cite{dragulesco3} concerning the cumulative probability of
income  in the United Kingdom and the United States, during the period 1994-1999 the United Kingdom was
characterized by a Pareto exponent in the interval $1.8<\alpha <2.3$ and
during the year 1998 the United States was characterized by a Pareto
exponent $\alpha \approx 1.7$. By assuming that the  cumulative wealth distributions have Pareto exponents similar to that of the correspondent income distributions, these different values of the Pareto exponent would imply that wealth is more concentrated in the United States than it is in
the United Kingdom. By supposing similar investment economies of the two countries, the higher value of the UK Pareto's index   can be interpreted in the context of our model
as being due to the social policies adopted in the UK, yielding a higher
social index $f$.

The inset in Fig. 6 shows that in the wealth interval [10,100] the
cumulative probability $P(w),$ for an ideal economy with $f=0.3$, $h=0.05$
and $r=0.075,$ can be fitted by an exponential function. A similar result is
found by Dr\u{a}gulescu and Yakovenko \cite{dragulesco3} in empirical
cumulative probability of wealth  in the United Kingdom. However, the exponential fit, that  would justify the application of the
Maxwell-Boltzmann distribution model, see Sec. 2, to the wealth as well as income distribution at low
value, is here only apparent and not real. These authors recognize that at low income, and we suppose at low wealth as well, the pdf is not monotonic because it increases and then decreases, and may be  better fit   by a Gamma distribution, but Dr\u{a}gulescu et
al. \cite{dragulesco3} do not have any model to reproduce this  behavior of the empirical data.  Instead, as Fig. 5 shows, this non-monotonic behavior of the distribution at low and middle wealth value is well reproduced by the trade-investment model. Therefore, caution must be exercised in drawing conclusion from the data fitting of the cumulative distribution alone.

\section{Wealth and income distributions in societies}

As anticipated in the introduction and in the previous section the full range empirical distributions of income of several countries \cite{dragulesco3,souma,montroll2,badger} present a  differentiation between the rich and  non-rich classes. Fig. 1 shows that, at least for United Kingdom, wealth and income distributions look very similar, therefore we suppose that such a  differentiation is a general property for the wealth distributions as well.    
The middle-low wealth range appears to follow a distribution that Dr\u{a}gulescu et
al. \cite{dragulesco3} fit with an exponential while Souma,
following Gibrat \cite{gibrat}, fits with a log-normal distribution. The higher wealth range, that involves only 1-2\% of the population, maintains a Pareto tail. This particular anomalous shape may be recovered by the trade-investment model as Figs. 7 and 8 show. 

A possible interpretation of the separation effect is that the
investment index varies among the members of society, as we have already
suggested in Eq. (\ref{firsteq6}). The simplest option is to assume that $N_{1}$
agents invest their wealth with a investment index $r_{1}$ and $N_{2}=N-N_{1}
$ agents invest their wealth with a different  investment index $r_{2}$: 
\begin{equation}
\frac{dW_{i}}{dt}=r_{1}~\xi ~W_{i}+\sum_{j=1(\neq i)}^{N}w_{ij} ~~~~,~~~ i\leq N_{1}  
\label{eqneqwt1} 
\end{equation}
\begin{equation}
\frac{dW_{i}}{dt}=r_{2}~\xi ~W_{i}+\sum_{j=1(\neq i)}^{N}w_{ij} ~~~~,~~~ i>N_{1} ~.
 \label{eqneqwt2}
\end{equation}
In the eventuality that the system is characterized by more than two
investment indexes, we have simply to expand the set of equations to
accommodate them. In theory, the investment index $r$ may change for each
trader.
Fig. 7 shows two cumulative distributions of our ideal economic society of
100,000 agents. The social index is $f=0.3$, the poverty index is $h=0.05$
and the investment index is $r_{1}=0.075$ or $r_{1}=0.055$ for 50,000 agents
and $r_{2}=0$ for the other 50,000 agents. The curves look very similar to the empirical distributions found in  Refs. \cite{dragulesco3} and  \cite{souma}, and in Fig. 1. The Pareto exponents are $\alpha =1.5\pm 0.02$ and $\alpha =2.5\pm
0.02$,  which are in the range the empirical distributions.

However, Fig. 7 shows that the Pareto law characterize only 1\% of the
population, while in the simulation we assumed that 50\% of the population
is characterized by a trade-investment economy and the other 50\% by a trade-alone economy. This fact suggests that
the multiplicative process of the trade-investment economy is effective  only for a small portion of the
population. In fact, Fig. 8 suggests a slightly different and deeper
interpretation of the dynamics of the trade-investment model. Fig. 8 shows
the cumulative probability of two trade-investment economies with two
different poverty index $h=0.03$ and $h=0.09$. The investment index $r=0.075$
and the social index $f=0.3$ are the same for all members of the society, so
the computer simulation is done by using only one kind of evolution
equation Eq. (\ref{firsteq6}). 

In fact, the explanation of Fig. 8 is quiet suggestive. The society may be
divided in two groups: the small rich class, almost 1\% of the population,
and all the others. The difference between the two groups is in the economic
mechanism that has the greatest effect on the wealth of each group. The
meaning of the results shown in Figs. 7 and 8 is that the change of wealth
of the rich class is more influenced by the investment economy while the
wealth of all the others is more influenced by the trade economy. In fact,
for people that are not rich the amount of wealth that they may gain or lose
with some investment is compatible or lower than what they may gain or lose
in trades. Instead, for the rich, trade alone may move only a small portion
of their wealth because most of their trades involve agents poorer than
themselves. So, for the rich, the investment part of the economy dominates,
and because the investment economy is described by a multiplicative process,
the wealth of the rich follows the Pareto inverse power law. For the other
side, by increasing the poverty index, the capacity of the trade to move
wealth is amplified and this explains how by increasing the poverty index $h,
$ the difference between the rich (that are characterized by the investment
economy that generates a Pareto's wealth distribution), and all the others (that are characterized by the trade economy  that generates a Gamma-like wealth distribution)
becomes more prominent, as shown in Fig. 8.

In summary, the distribution of wealth shows an anomalous shape well described by a Gamma distribution at low and middle wealth and a Pareto's tail at high wealth. This complex shape  may be approximately described by
\begin{equation}
p(w)=a~w^{\gamma }\left[ \exp(-b~w)+\frac{1}{(1+c~w)^{\gamma +\delta }}\right]~.
 \label{fitt3}
\end{equation}
that is a pdf of the type suggested by Montroll and Shlesinger, Eq. (\ref{renorm1}), and depends on
many parameters that characterize the difference between the trade and the
investment economy. The Pareto index increases by increasing the social
index $f$ or by decreasing the investment index $r$. The dependency of the
wealth condensation phenomenon on the poverty index $h$ seems more complex. In
fact, in Sec. III we determined that wealth condensation in a trade-alone
economy increases by increasing the poverty index $h$. However, Fig. 8 shows
that in a trade-investment economy, wealth condensation may also decrease by
increasing the poverty index $h$. Therefore, wealth condensation is a
non-monotonic function of $h$. We stress that it is necessary to have a
valid trade-interaction model to obtain all the above results. We have shown
that by changing the value of the three indexes, our trade-interaction model
is able to give a consistent explanation of the observed 
distribution of wealth.

\section{Conclusion}

While some linear asset exchange models like those based on the mean-field and  kinetic theory approximations may present serious economical paradoxes, the science of complexity and physically-based models  may still yield possible explanations of the distribution of wealth in a society. The nonlinear trade-investment model determines the main mechanisms necessary
to obtain a stable three-class society with a Pareto distribution for high
incomes. The mechanisms present in the trade-investment model are an asymmetric trade
interaction that statistically favors the less wealthy of the two traders
involved in the trade and a multiplicative stochastic investment process. A
symmetric trade-alone mechanism would yield a strong wealth condensation phenomenon
and therefore an unstable society. Both  mechanisms are necessary to overcome the wealth condensation effect and stabilize the society. The detailed nature of the relative
contribution of the two mechanisms are determined by the three control
parameters: the social index \textit{f}, the investment index \textit{r} and
the poverty index \textit{h}.

Here, we cannot do a detailed economic analysis of the mechanisms that
advantage the poor in society by inhibiting the
transfer of the entire wealth to the rich. In general,  the richer party is   less risk averse
when bargaining over a given amount than is the poorer party, therefore the poorer party should be a stronger bargainer than the rich to get the better of the
deal. In particular, we can consider the tendency of the poor to look for the best price
opportunity for most single items for saving money, the tendency of the prices to increase in places where the customers are supposed wealthy, the  employee rights that
advantage workers by mean of adequate salary policies,  the graduated income and luxury tax
policies that require the rich to pay a higher percentage of taxes than is
paid by the wage earner,  and many others.  In the absence of such
social and psychological pro-poor mechanisms it is easy to envision that the society would collapse because the
entire wealth  would condense into the hands of a very few
people. Also we observe that the above mechanisms can be interpreted as a kind of generalization of the  {\it price discrimination in monopoly} theory \cite{pricedisc} applied to the social classes of an entire society.  

Investment is the dominant economic tool of the very rich. Montroll and
Shlesinger \cite{montroll2} argued that the rich have economic mechanisms
available to them that the ordinary citizen does not, and these mechanisms
provide the amplification that generates the inverse power law distribution.
The trade-investment model  incorporates this amplification  through the multiplicative form of the investment term in the
dynamic equation. The magnitude of the amplification factor is determined by
the investment index \textit{r}.

Trade is the dominant economic tool of the majority of society. Salaries and taxes are considered particular trades. In the upper
economic limit of the middle class some individuals increase their wealth as
much through investment as they do through trade, providing a transition  from a Gamma-like distribution that characterizes a trade-dominated economy to a Pareto's tail distribution produced by the investment economy.  What distinguishes the
trade-investment model from other models that have been proposed is the
recognition that in order for a society to be stable, rather than the poor
being exploited in trades with the rich they must have an advantage, at
least in a statistical sense. This statistical advantage acts to narrow the
gap in wealth between the richest and poorest members of society. The two
parameters that control this narrowing of the gap is the social index $f$,
that enables a poor individual to optimize his/her return in a trade with
the rich, and the poverty index $h$, that determines the fraction of the
poorer agent's wealth that can be transferred in any given transaction.

When we implement the trade-investment model with the three indices $f$, $h$
and $r$, the wealth pdf assumes the inverse power-law distribution of Pareto
at the high wealth end, and, more realistically, still retains a small but
finite population at the low wealth end. Moreover, the cumulative
distribution of wealth assumes the shapes depicted in Figs. 7 and 8 that are similar to the shape depicted in Fig. 1, in
which the rich class is distinguished by following Pareto's inverse power
law and the cumulative distribution of wealth for the remainder of society
is apparently well fit by  a Gamma distribution. Numerical simulations  shown that the dynamics of the
trade-investment model Eq. (\ref{firsteq6}) reproduces the shape of the
universal structure of the cumulative distribution of wealth that is assumed to be similar to the cumulative distributions of income obtained for
phenomenological data from around the world.

{\ {\large \textbf{Acknowledgment:}}}\newline
N.S. thanks the Army Research Office for support under grant DAAG5598D0002. We would like to thank prof. M. Boianovsky and prof. T. Groves for some useful discussion.

\newpage
\begin{table}
\begin{tabular}{|c|c|c|c|c|}
\hline
        & a & d & $c$ & $\eta$  \\ \hline
f=0.2 & 3e-3 $\pm$ 3e-4 &  0.147 $\pm$ 2e-3 & 1.64 $\pm$ 0.05 & 1.9 $\pm$ 0.05  \\ \hline
f=0.3 & 5e-3 $\pm$ 1e-3 & 0.56 $\pm$ 0.02 & 0.8 $\pm$ 0.1 & 3.9 $\pm$ 0.2 \\ \hline
f=0.5 & 0.10 $\pm$ 0.06 & 1.9 $\pm$ 0.06 & 0.5 $\pm$ 0.1 & 6.8 $\pm$ 0.5 \\ \hline
\end{tabular}
\caption{Fitting parameters of Eq. (\ref{fitt1}) for the trade-alone economy (r=0), see Fig. 3. The poverty index is fixed $h=0.05$.}
\end{table}

\begin{table}
\begin{tabular}{|c|c|c|c|c|}
\hline
         & a & b & $\gamma$ & $\delta$ \\ \hline
f1(w) & 6e-5 $\pm$ 1e-5 & 2.6e-2 $\pm$ 2e-3 & 2 $\pm$ 0.1 & 2.15 $\pm$ 0.05 \\ \hline
f3(w) & 5e-4 $\pm$ 1e-4 & 0.17 $\pm$ 0.02 & 5.8 $\pm$ 0.6 & 3.5 $\pm$ 0.1 \\ \hline
\end{tabular}
\caption{Fitting parameters of Eq. (\ref{fitt2}) for the Trade-investment economy, see Fig. 5. The social and poverty index are fixed; $f=0.3$ and $h=0.05$.}
\end{table}

%\onecolumn
\newpage
\begin{figure}
\epsfig{file=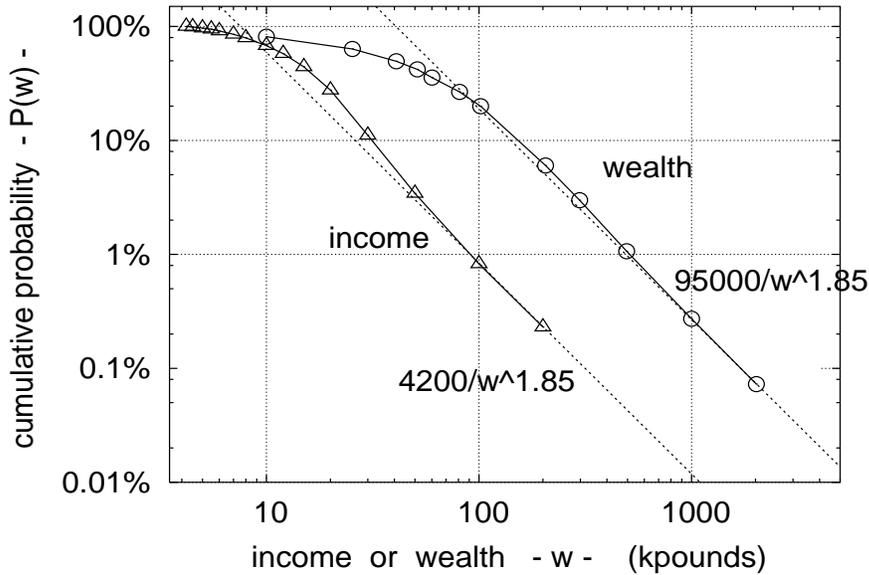,height=12cm,width=8cm,angle=-90}
\caption{Cumulative wealth (1996) and income (1998-1999) distributions in United Kingdom. }
\end{figure}

\newpage
\begin{figure}
\epsfig{file=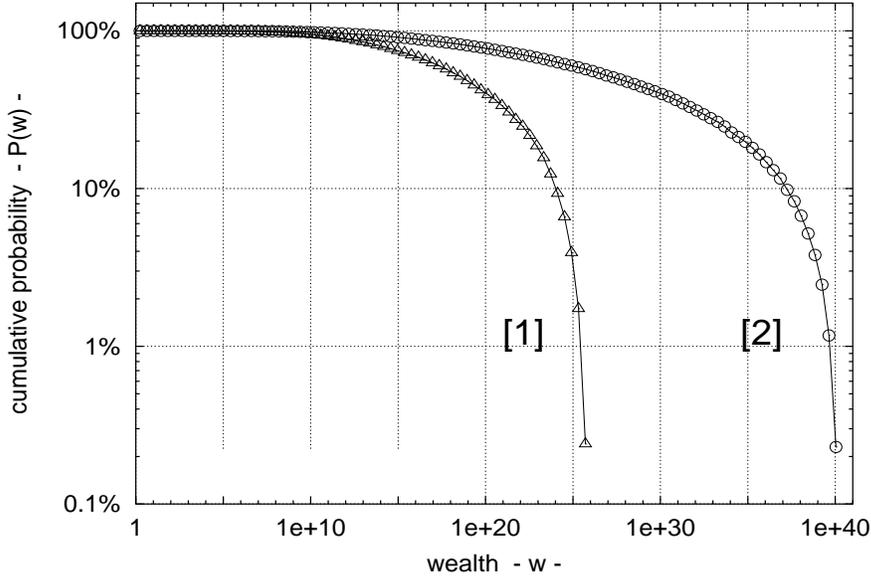,height=12cm,width=8cm,angle=-90}
\caption{Cumulative wealth distribution for the symmetry chance model. The indexes are: $h=0.05$, $f=0$ and $r=0$. Case [1] is after 100 million trade-interactions, and case [2] is after 200 million trade-interactions. The initial wealth distribution is uniform. The figure shows that this model yields to a huge wealth gap between the rich and poor that increases with the number of interactions.   The wealth is measured in units of the poorest agent's wealth. }
\end{figure}

\newpage
\begin{figure}
\epsfig{file=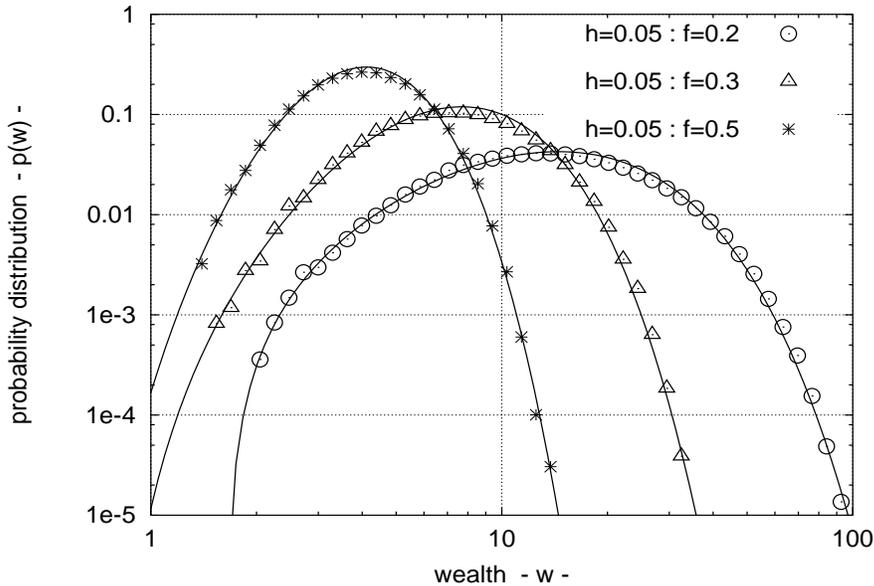,height=12cm,width=8cm,angle=-90}
\caption{Wealth probability density for the asymmetric chance model with a fixed  poverty index  $h=0.05$.    The wealth condensation increases by decreasing the social index $f$. The investment index is $r=0$. The distributions are fitted by a Gamma distribution Eq. (\ref{fitt1}). The fitting parameters are in Table I. The wealth is in units of the poorest agent's wealth.  }
\end{figure}

\newpage
\begin{figure}
\epsfig{file=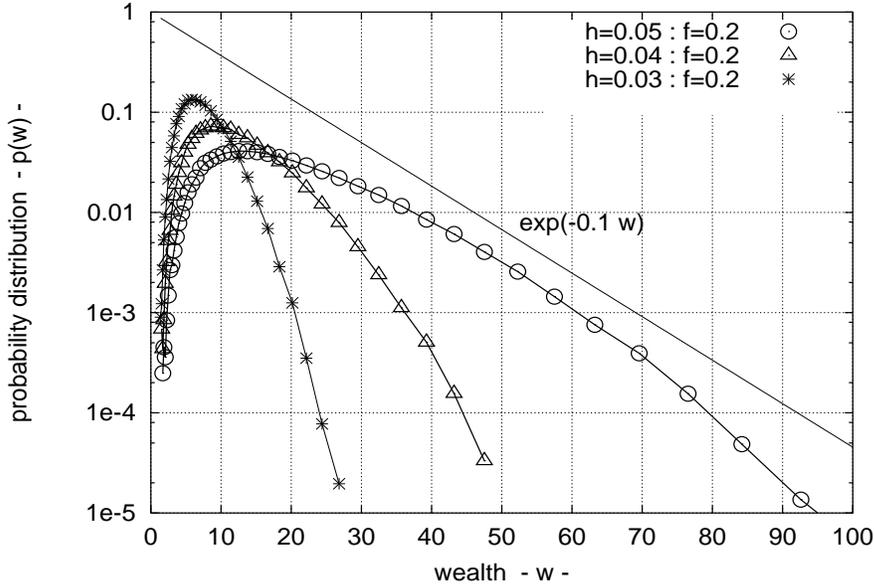,height=12cm,width=8cm,angle=-90}
\caption{Wealth probability density for the asymmetric chance model with a fixed social index $f=0.2$.   The wealth condensation increases by increasing the poverty index $h$. The distributions are compared to an exponential Maxwell-Boltzmann distribution. The
wealth is in units of the poorest agent's wealth.  }
\end{figure}

\newpage
\begin{figure}
\epsfig{file=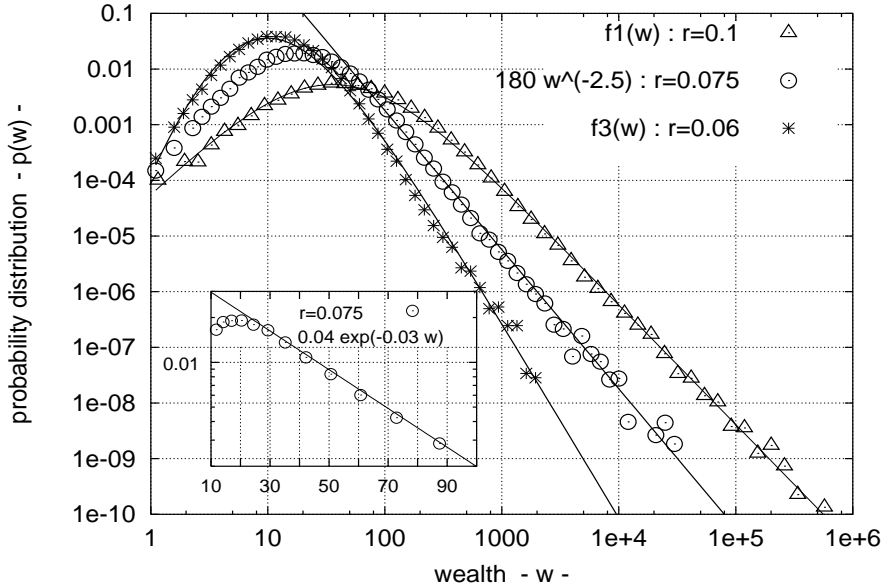,height=12cm,width=8cm,angle=-90}
\caption{Trade-investment economy. The social and poverty index are fixed; $f=0.3$ and $h=0.05$. The probability distributions (tringles) and  (stars) are fitted by using Eq. (\ref{fitt2}). The fitting parameters are in Table II. The tail of the probability distribution  (circles) is fitted  by a power law of the type $1/x^{\mu+1}$ where $\mu=1.5$ in the Pareto's exponent.  The small picture shows the probability distribution  (circles) in the interval [10:100] that is fitted with an exponential Maxwell-Boltzmann distribution. The wealth is measured in units
of the poorest agent's wealth. }
\end{figure}

\newpage
\begin{figure}
\epsfig{file=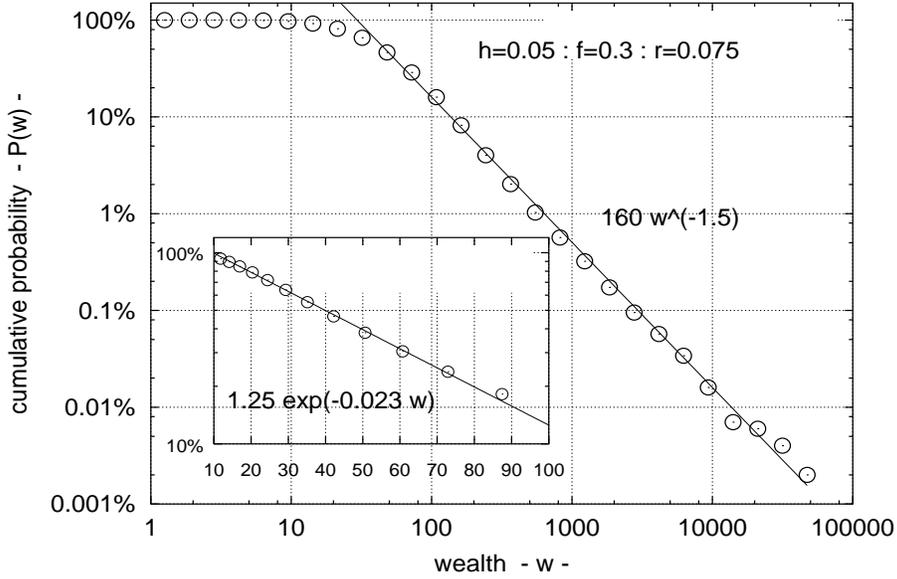,height=12cm,width=8cm,angle=-90}
\caption{Cumulative probability for an trade-investment economy with $h=0.05$, $f=0.3$ and $r=0.075$. The Pareto's exponent is $\mu=1.5\pm0.02$. The little picture shows that in the interval [10:100] the P(w) can be apparently fitted by an exponential Maxwell-Boltzmann distribution. The wealth
is measured in units of the poorest agent's wealth.}
\end{figure}

\newpage
\begin{figure}
\epsfig{file=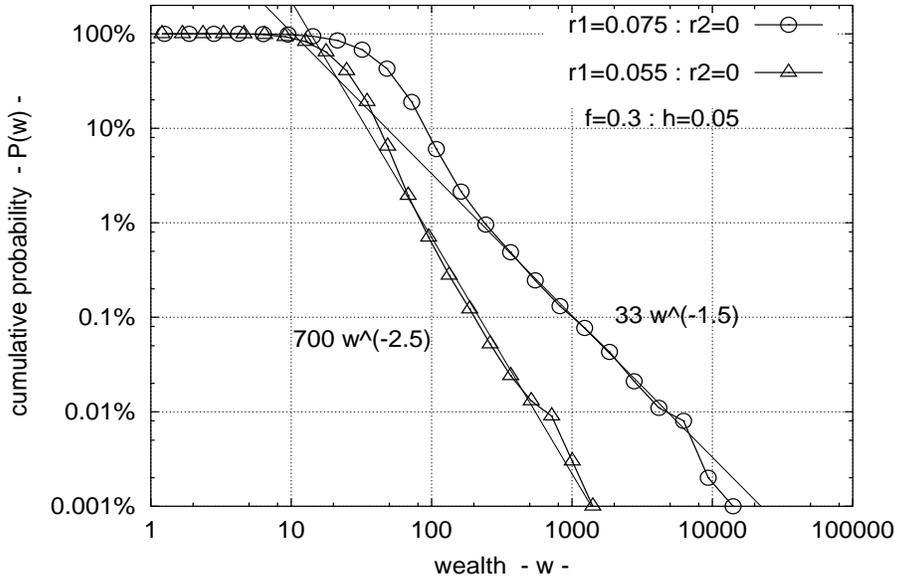,height=12cm,width=8cm,angle=-90}
\caption{Cumulative probability for a double trade-investment economy. One half of the population has the investment index $r_2=0$, the other half of the population has in one case $r_1=0.075$ and in the other $r_1=0.055$. In both
cases the social index is $f=0.3$ and the poverty index is $h=0.05$. The Pareto's exponents are $\mu=1.5\pm0.02$ and $\mu=2.5\pm0.02$.  The wealth is measured in units of the poorest agent's wealth.}
\end{figure}

\newpage
\begin{figure}
\epsfig{file=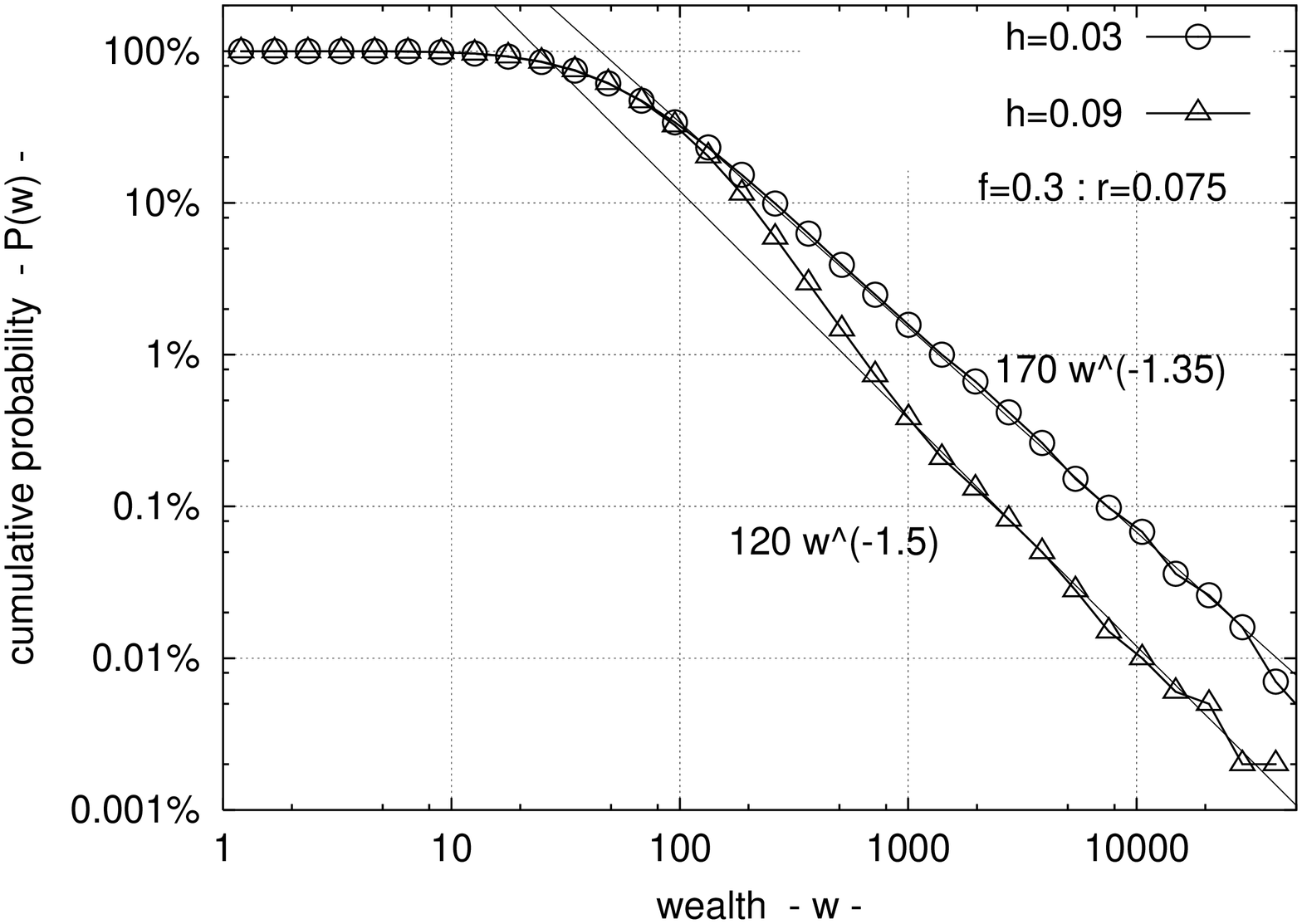,height=12cm,width=8cm,angle=-90}
\caption{Cumulative probability for two  trade-investment economies with different poverty index $h$.   In both
cases the social index is $f=0.3$ and the investment index is $r=0.075$. The Pareto's exponents are $\mu=1.5\pm0.02$ and $\mu=1.35\pm0.02$.   The wealth is measured in units of the poorest agent's wealth.}
\end{figure}


\begin{thebibliography}{99}
\bibitem{west1}  B.J. West, \textit{Physiology, Promiscuity and Prophecy at
the Millenium: A Tale of Tails}, World Scientific, Singapore (1999).



\bibitem{pareto} V. Pareto, {\it Manuale di economia politica,} Milano, Societ\`a Editrice, 1906. V. Pareto, \textit{Cours d'Economie Politique}, Lausanne and Paris (1897).

\bibitem{black} J. Black, {\it Dictionary of Economics}, Oxford University Press, New York (2002). 

\bibitem{dragulesco3}  A. Dr\u{a}gulescu and V. M. Yakovenko, \textit{%
``Exponential and power-law probability distributions of wealth and income
in the United Kingdom and the United States,''} Physica A, \textbf{299},
213-221 (2001).

\bibitem{souma}  W. Souma, \textit{``Universal structure of the personal
income distribution,''} Fractals, \textbf{9} No. 4, 463-470 (2001).

\bibitem{stati}{\it http://www.inlandrevenue.gov.uk/stats/} : Inland Revenue, National Statisics, UK.



\bibitem{stanley2}  H. E. Stanley and R. N. Mantegna, in \textit{An
introduction to econophysics}, Cambridge university press, Cambridge UK
(2000).



\bibitem{montroll2}  E.W. Montroll and M.F. Shlesinger, {\it ``Maximum-entropy formalism, fractals, scaling phenomena, and 1/f noise- a tale of tails,"}
J. Stat. Phys.
\textbf{32}, 209 (1983).

\bibitem{stanley1}  H.E. Stanley, \textit{Introduction to Phase Transitions
and Critical Phenomena}, Oxford University Press, New York (1971).

\bibitem{list}  B. Gutenberg and C.F. Richter, {\it ``Frequency of earthquakes in California"} Bull. Seismol. Soc. Am.
\textbf{34}, 184 (1944); J.M. Carlson, J.S. Langer and B.E. Shaw, {\it ``Dynamics of earthquake faults,"} Rev. Mod. Phys. \textbf{66}, 657 (1994).
 O. Peters, C. Hertlein and K. Christensen, {\it ``A complexity view of rainfall,"}
 Phys. Rev. Lett. \textbf{88}, 018701 (2002).
  C.K. Peng, J. Mietus, J.M. Hausdorff, S. Havlin, H.G.
Stanley and A.L. Goldberger, {\it ``Long-range anticorrelations and non-gaussian behavior of the heartbeat,"} Phys. Rev. Lett. \textbf{70}, 1343 (1993).
  B.J. West, R. Zhang, A.W. Sanders, S. Miliyar, J. H.
Zucherman and B. Levine, {\it ``Fractal fluctuations in cardiac time series,"} Physica A \textbf{270}, 552 (1999).
  B.J. West, V. Bhargava and A.L. Goldberger, {\it ``Beyond the Principle of Similtude: Renormalization in the Bronchial Tree"}  J. Appl.
Physiol. \textbf{60} 3, 1089-1097 (1986).
  P. Grigolini, D. Leddon, N. Scafetta, \textit{``The
Diffusion entropy and waiting time statistics of hard x-ray solar flares,''}
Phys. Rev. E \textbf{65}, 046203 (2002).
  M.F. Shlesinger, J. Klafter and B.J. West, {\it ``L\'evy dynamics of enhanced diffusion-application to turbulence,"} Phys. Rev.
Lett. \textbf{58}, 1100 (1987).

\bibitem{bouchaud1}  J. P. Bouchaud and M. M\'{e}zard, \textit{``Wealth
condensation in simple model of economy,''} Physica A \textbf{282}, 536-545
(2000).



\bibitem{dragulesco1}  A. Dr\u{a}gulescu and V. M. Yakovenko, \textit{%
``Statistical mechanics of money,''} Eur. Phys. J. \textbf{%
B17}, 723-729, (2000).

\bibitem{dragulesco2}  A. Dr\u{a}gulescu and V. M. Yakovenko, \textit{%
``Evidence for the exponential distribution of income in the USA,''} Eur. Phys. J. \textbf{B20}, 585-589, (2001).

\bibitem{Chakraborti}  A. Chakraborti, B.K. Chakrabarti, \textit{%
``Statistical mechanics of money: how saving propensity affects its
distribution,''} Eur. Phys. J.  \textbf{B17}, 167 (2000).



\bibitem{ispolatov}  S. Ispolatov, P. L. Krapivsky and S. Redner, \textit{%
``Wealth distribution in asset exchange models,''} Eur. Phys. J. \textbf{B2}, 267-276, (1998).



\bibitem{malcai}  O. Malcai, O. Biham, P. Richmond and S. Solomon, \textit{%
``Theoretical analysis and simulations of the generalized Lotka-Volterra
model,''} Phys. Rev. E \textbf{66}, 031102 (2002).

\bibitem{burda}  Z. Burda, D. Johnston, J. Jurkiewicz, M. Kaminski, M. A.
Nowak, G. Papp and I. Zahed, \textit{``Wealth condensation in pareto
macroeconomics,''} Phys. Rev. E \textbf{65}, 026102 (2002).

\bibitem{levy}  M. Levy, \textit{``Market Efficiency, the Pareto Wealth
Distribution, and the Levy Distribution of Stock Returns,''} in The Economy
as an Evolving Complex System III, S. Durlauf and L. Blume (Eds.), Oxford
University Press, Oxford forthcoming.



\bibitem{Boucha}  J. P. Bouchaud, \textit{``Power laws in economics and
finance: some ideas from physics,''} Quantitative Finance \textbf{1}, No 1
(January 2001) 105-112.

\bibitem{sornette}  D. Sornette, \textit{``Fokker-Planck equation of
distributions of financial returns and power laws,''} Physica A \textbf{290}
(1-2), 211-217 (2001).

\bibitem{solomon}  Z.-F. Huang and S. Solomon, \textit{``Stochastic
Multiplicative Processes for Financial Markets,''} Physica A \textbf{306}
412-422 (2002).

\bibitem{lindenberg}  K. Lindenberg and B.J. West, \textit{The
Nonequilibrium Statistical Mechanics of Open and Closed Systems}, VCH
Publishers, New York (1990).

\bibitem{lawoneprice}  An economic rule which states that in an efficient
market, a security must have a single price, no matter how that security is
created. For example, if an option can be created using two different sets
of underlying securities, then the total price for each would be the same or
else an arbitrage opportunity would exist. http://www.investorwords.com .
About the deviation from the law of one price, see also: K.A. Froot, {\it ``The
Law of One Price over 700 Years,"} http://www.hbs.edu, unpublished.

\bibitem{mauro}  M. Boianovsky and V.J. Tarascio, {\it ``Mechanical inertia and
economic dynamics: Pareto on business cycles,"} J. of the History of Economic
Thought, \textbf{20} 5 (1998).


\bibitem{feller}  W. Feller, \textit{An Introduction to Probability Theory
and Its Applications}, 3rd Edition, Vol.1, John Wiley \& Sons, New York
(1968).

\bibitem{tsallis}  C. Tsallis, \textit{``Possible generalization of
Boltzmann-Gibbs statistics,''} J. Stat. Phys. \textbf{52}, 479 (1988).

\bibitem{montroll}  E.W. Montroll and W.W. Badger, \textit{Introduction to
the Quantitative Aspects of Social Phenomena}, Gordon \& Breach, New York
(1974).

\bibitem{badger}  W.W. Badger, in \textit{Mathematical Models as a Tool for
the Social Sciences}, Editor B.J. West, pp.87-120, Gordon \& Breach, New
York (1980).

\bibitem{gibrat}  R. Gibrat, \textit{Les In\'{e}galit\'{s} \'{E}conomique},
Sirey, Paris (1931).

\bibitem{pricedisc} J. M. Henderson and R. E. Quandt, {\it Microeconomic theory: a mathematical approach,} McGraw-Hill, USA (1971).

\end{thebibliography}
\end{document}